\documentclass[11pt]{article}
\usepackage{hyperref}

\usepackage{graphicx}
\usepackage{bm}
\usepackage{multirow}
\usepackage{subfigure}
\usepackage{fancyhdr}

\usepackage[T1]{fontenc} 
                   
\usepackage{amssymb}
\usepackage{amsmath} 

\pdfoutput=1

\begin{document}
\title{Vortex Shedding From a \\ Flexible Hydrofoil}
\author{Matthieu Dreyer, Mohamed Farhat \\
\\\vspace{6pt} \'Ecole Polytechnique Fédérale de Lausanne}
\maketitle


When a solid body is placed in a fluid stream it can exhibit a separated flow that extends to its wake. Above a critical value of Reynolds number, the detachment of the boundary layer on both upper and lower surfaces forms two shear layers, which generate a periodic array of discrete and alternate vortices termed Von Kármán street. The body experiences a fluctuating lifting force caused by the asymmetric formation of vortices. When the shedding frequency approaches a resonance frequency of the combined fluid-structure system, the coherence of the vortex shedding is considerably enhanced leading to a significant increase of the vibration amplitude. \\

The present fluid dynamics video illustrates the vortex shedding in the wake of a Naca0009 hydrofoil having 100 mm chord length. The hydrofoil is made of polyoxymethylene type C (POM C) with an elasticity modulus of 3.1 GPa. The resulting amplitude of flow induced vibrations at resonance is in the millimeter range, which is comparable to the boundary layer thickness at the trailing edge. The tests were carried out in the EPFL high-speed cavitation tunnel, Avellan et al. [1]. The flow velocity is tuned to excite one of the following mode shapes: the 2$^{nd}$ torsion mode (530 Hz), the 3$^{rd}$ bending mode (700 Hz), and the 3$^{rd}$ torsion mode (1030 Hz). The corresponding flow velocities are respectively 9.5, 12.5 and 17 m/s. Cavitation, which develops in the core of alternate vortices, is used as a tracer for flow visualization. Although the cavitation is known to produce an increase of the shedding frequency, the wake structure remains unaffected, Ausoni et al. [2]. \\

\vspace{0.5cm} 
A complex fluid structure interaction is revealed by the present video. For all tested mode shapes, the process of vortex shedding remains periodic. The Strouhal number based on the trailing edge thickness is around 0.17, which is significantly lower than similar rigid hydrofoil [2]. If the vibration amplitude is added to the trailing edge thickness for the length scale, the Strouhal number perfectly matches the one of rigid hydrofoil. The flow visualization clearly illustrates how the alternate shedding prevails despite the large vibration amplitudes. However, the cavitation occurrence reveals how vorticity starts building on antinodes at their maximum amplitude for the upper surface and minimum amplitude for the lower surface. As a result, the vorticity lines are strongly deformed as they gradually leave the trailing edge. Nevertheless, a parallel and alternate shedding is quickly recovered in the wake, in line with stability conditions stated by Von Kármán theory. 

\begin{figure}[!h]
\begin{center}
\includegraphics[width=1\textwidth]{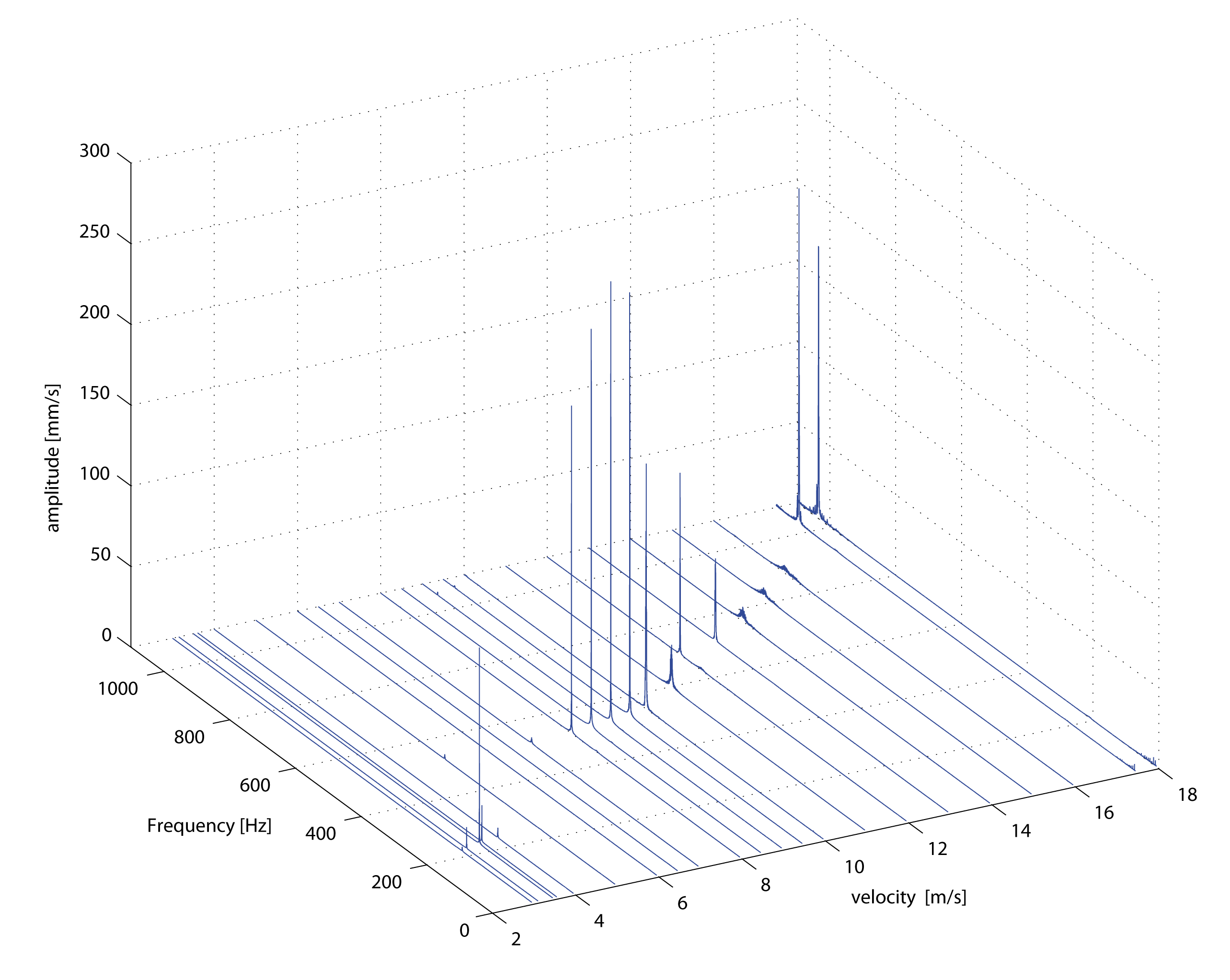}
\caption{ Waterfall spectra of the vortex-induced vibration signals,for different free-stream velocities. Cavitation free regime.}
\label{fig_fft}
\end{center}
\end{figure}

\end{document}